# Broadband coherent Raman scattering spectroscopy at 50,000,000 spectra/s

Takuma Nakamura[1], Kazuki Hashimoto[1], and Takuro Ideguchi[1,*]

[1]Institute for Photon Science and Technology, The University of Tokyo, Tokyo 113-0033, Japan
*ideguchi@ipst.s.u-tokyo.ac.jp

**Abstract**

Raman scattering spectroscopy is widely used as an analytical technique in various fields, but its measurement process tends to be slow due to the low scattering cross-section. In the last decade, various broadband coherent Raman scattering spectroscopy techniques have been developed to address this limitation, achieving a measurement rate of about 100 kSpectra/s. Here, we present a significantly increased measurement rate of 50 MSpectra/s, which is 500 times higher than the previous state-of-the-art, by developing time-stretch coherent Raman scattering spectroscopy. Our newly-developed system, based on a mode-locked Yb fiber laser, enables highly-efficient broadband excitation of molecular vibrations via impulsive stimulated Raman scattering with an ultrashort femtosecond pulse and sensitive time-stretch detection with a picosecond probe pulse at a high repetition rate of the laser. As a proof-of-concept demonstration, we measure broadband coherent Stokes Raman scattering spectra of organic compounds covering the molecular fingerprint region from 200 to 1,200 $cm^{-1}$. This high-speed broadband vibrational spectroscopy technique holds promise for unprecedented measurements of sub-microsecond dynamics of irreversible phenomena and extremely high-throughput measurements.

**Introduction**

Raman scattering spectroscopy provides detailed information about molecular vibrations used for chemical analyses in various fields. The traditional method of spontaneous Raman scattering spectroscopy offers broadband spectra with a high spectral resolution, allowing for detailed analysis of multiple vibrational lines, but its small scattering cross-section results in a long measurement time. Coherent Raman scattering (CRS) spectroscopy, such as coherent anti-Stokes Raman scattering (CARS) and stimulated Raman scattering (SRS) spectroscopy, provides a significant signal enhancement and increases the spectral measurement rate[1,2]. In recent years, various high-speed broadband CRS spectroscopy techniques have been developed, including dual-comb CARS[3], broadband CARS with a dispersive spectrometer[4], frequency-swept time-encoded SRS[5], rapid-scan Fourier-transform CARS[6], rapid-scan chirped-pulse SRS[7], and photonic time-stretch SRS[8]. Since 2016, when rapid-scan Fourier-transform CARS demonstrated a measurement rate higher than 10 kSpectra/s for the first time[6], other updated systems using the aforementioned techniques have also shown improved spectral acquisition rates. Despite the efforts, the measurement rates of those systems have remained in the range of around 10-100 kSpectra/s, with the highest reported rate being 111 kSpectra/s[9]. Representative achievements are summarized in Table S1 of Supplementary Note 1. These high-speed techniques have been applied to various applications, such as label-free vibrational hyperspectral imaging[10] and flow cytometry[11–13]. Since the spectral measurement rate limits the imaging speed and throughput, there is a high demand for faster measurement techniques to advance this field.

Among various high-speed broadband spectroscopy techniques, time-stretch spectroscopy[14], also known as dispersive Fourier-transform spectroscopy, has the advantage of achieving a high spectral measurement rate with a high signal-to-noise ratio (SNR). In this technique, a femtosecond pulse is temporally stretched to nanoseconds by adding a significant amount of dispersion to the pulse, resulting in time-to-wavelength conversion. It enables ultrafast and continuous spectral sampling of a pulse train in the time domain using a high-speed photodetector and a digitizer. Time-stretch spectroscopy has been applied to broadband linear absorption spectroscopy, resulting in spectral acquisition rates of several tens of MSpectra/s, first in the near-infrared region[15] and, more recently, in the mid-infrared region[16,17]. Time-stretch CRS spectroscopy has also been demonstrated, with the first demonstration for measuring Raman amplification in a silicon waveguide in the telecom region[18]. While this system showed ultrafast dynamics of gain switching at a temporal resolution of 40 ns, it could only be used for measurements where the sample has a huge Raman gain. The narrow spectral bandwidth of 35 cm$^{-1}$ is another limitation of this system. As another demonstration, broadband time-stretch CARS and coherent Stokes Raman scattering (CSRS) spectroscopy of gas phase samples was reported[19]. Although it showed a single-shot acquisition time of 20 ns, its spectral measurement rate was limited to 20 Spectra/s due to the use of an amplified femtosecond laser source with an extremely large excitation pulse energy of 30 mJ for gas phase spectroscopy. Recently, time-stretch SRS was demonstrated for measuring liquid-phase organic molecules with an acquisition rate of 20 kSpectra/s[8]. However, this system also used an amplified femtosecond laser at a low repetition rate of 80 kHz, about three orders of magnitude lower than those of mode-locked oscillators. As time-stretch spectroscopy can, in principle, be operated at tens of MSpectra/s by utilizing a high repetition rate of a mode-locked laser, there is significant room for improvement in the spectral measurement rate.

In this work, we present time-stretch coherent Stokes Raman scattering (TS-CSRS) spectroscopy that provides broadband and high-resolution spectra at a remarkable spectral measurement rate of 50 MSpectra/s. Our approach involves the use of an ultrashort femtosecond pulse to efficiently excite molecular vibrations via impulsive stimulated Raman scattering (ISRS)[20] and the sensitive time-stretch detection of a CSRS spectrum using a picosecond probe pulse. To demonstrate this technique, we developed a passively synchronized dual-color femtosecond-picosecond laser system based on a Yb:fiber mode-locked laser. It enables broadband ISRS excitation in the range of 200 to 1,200 cm$^{-1}$, and continuous detection of time-stretch spectra at 50 MSpectra/s with moderate pulse energies in the order of nJ. To the best of our knowledge, the demonstrated measurement rate is about 500 times higher than the fastest broadband CRS spectroscopy. Also, it is the highest measurement rate in all the previously reported broadband vibrational spectroscopy techniques in the molecular fingerprint region, including mid-infrared linear absorption spectroscopy techniques.

## Results
**Principle of TS-CSRS spectroscopy.** Figure 1 shows a schematic and working principle of our TS-CSRS system. A detailed schematic is depicted in Fig. S1 of Supplementary Note 2. The system comprises a 50-MHz passively synchronized fs-ps two-color laser, broadband CSRS generation, and time-stretch detection of the CSRS spectra. This method involves impulsive excitation of molecular vibrations over a broad spectral range via the ISRS process with an ultrashort fs pulse[20]. The excitation pulse duration determines the measurable spectral bandwidth because all Raman-active molecular vibrations with cycles longer than the excitation pulse are simultaneously excited. Therefore, using ultrashort pulses at the level of 10 fs promises highly efficient excitation, enabling high-speed and broadband measurements. Then, the molecular vibrations are probed with a narrow-band ps pulse, which is phase-modulated due to the refractive index changes associated with the molecular vibrations. This process is similar to the three-color CARS[4], while we detect CSRS in this work. The red-shifted (Stokes) scattered photons are filtered out and temporally stretched in a long optical fiber for dispersive Fourier transformation[14]. Finally, the temporal intensity profile of the stretched pulse is recorded using a fast photodetector and an oscilloscope. Note that the time-stretch detection provides high-speed capability owing to a single photodetector while keeping a high SNR due to the nature of the frequency-swept detection[16,17]. The combination of the highly-efficient ISRS excitation and the highly-sensitive time-stretch detection enables the unprecedented measurement rate of broadband CRS spectroscopy. Detailed descriptions of each part of the system are provided below.

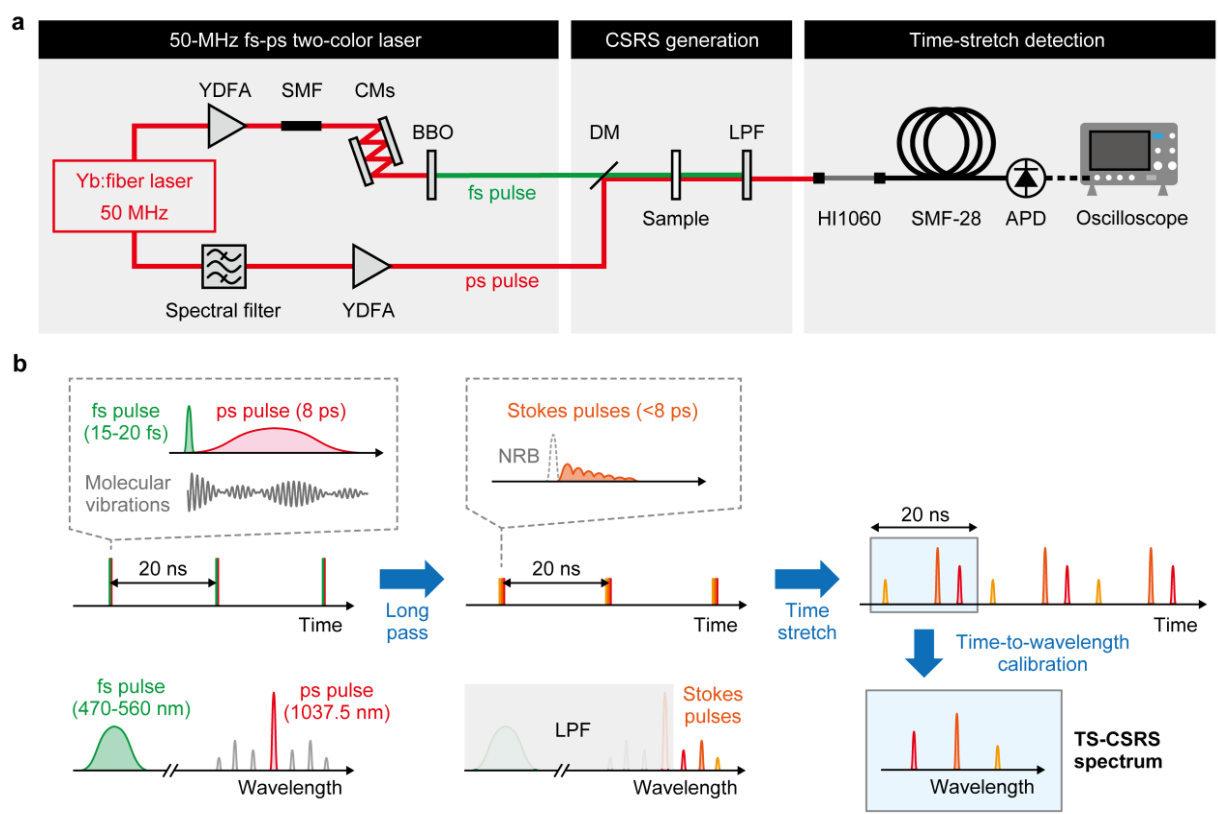

**Fig. 1** Time-stretch coherent Stokes Raman scattering (TS-CSRS) spectroscopy. **a,** A schematic diagram of TS-CSRS spectroscopy. YDFA: Yb-doped fiber amplifier, SMF: single-mode fiber, CMs: chirped mirrors, BBO: β-barium borate, DM: dichroic mirror, LPF: long-pass filter, APD: avalanche photodiode. **B,** Working principle of TS-CSRS spectroscopy. NRB: non-resonant background (omitted in the wavelength-domain picture for simplicity.)

**50-MHz passively-synchronized fs-ps two-color laser.** We developed a passively-synchronized dual-branch laser system for generating fs-excitation and ps-probe pulses. The system is based on a Yb:fiber mode-locked laser running at a repetition rate of 50 MHz. In a branch for generating fs-excitation pulses, seed pulses from the master laser are amplified with a polarization maintaining (PM) Yb-doped fiber amplifier (YDFA), spectrally broadened with a 4-mm single-mode fiber (HI1060), and temporally compressed using chirped mirrors. It generates 12-fs ultrashort pulses with a center wavelength of 1,030 nm and a pulse energy of 60 nJ. The details of the high-power and ultrashort pulse generation are described in our previous work[21]. The ultrashort pulses are focused onto a 50-μm β-barium borate (BBO) crystal for second harmonic generation (SHG), which generates 15-20 fs pulses with a spectrum spanning from 470 to 560 nm (see Fig. S1b of Supplementary Note 2). A thin BBO is used to minimize the effect of the group velocity mismatch in the crystal (100 fs/mm). The pulse energy of the SHG output is 3.0 nJ, and we use up to 1.6 nJ for the TS-CSRS measurements. Details of dispersion management of the fs SHG pulses are described in Methods.

In the other branch, seed pulses from the master laser are spectrally filtered by coupling a diffracted beam from a grating (1,200 lines/mm) into a single-mode fiber, generating ps-pulses with a bandwidth of 0.26 nm centered at 1,037.5 nm. These pulses are amplified with homemade two-stage PM-YDFAs. A large mode area fiber with a core diameter of 25 μm is used for the second amplifier for suppressing the undesired spectral broadening caused by self-phase modulation (SPM). The amplified spontaneous emission (ASE) is removed with a narrow bandpass filter. Figure S1c of Supplementary Note 2 shows the spectra of seed and amplified pulses. A slight SPM effect is observed after the second amplifier, resulting in a linewidth of <0.38 nm. The pulse duration should be about the transform limit of 8 ps as the dispersion added by the fiber amplifier is negligible. The pulse energy for CSRS spectroscopy can be adjusted by the pump power of the second amplifier up to 8.4 nJ. The polarization of the probe pulses is parallel to the pump pulses. The timing jitter of a two-branch laser is typically in the sub-fs level[22], which is negligible for the CSRS measurement with ps-probe pulses.

**Broadband CSRS generation.** The fs-excitation and ps-probe pulses are collinearly focused onto the sample in a 2-mm-long cuvette made of UV-grade fused silica. An off-axis parabolic mirror (OAPM) with a focal length of 15 mm is used instead of a conventional objective lens to avoid adding excess chromatic aberration and dispersion. The estimated effective numerical aperture (NA) is less than 0.15. The delay between the fs and ps pulses is adjusted using a free-space delay line. The excitation and probe pulse energies are 1.6 and 8.4 nJ, respectively. The ps-probe pulse generates both Stokes and anti-Stokes Raman scattering through refractive index modulations induced by molecular vibrations. We take the Stokes scattering, which lies up to 1,200 nm, by filtering out using a 1,050-nm long-pass filter because this spectral region gives a lower propagation loss in an optical fiber used in this work than that of anti-Stokes scattering. The dual-color configuration ensures the Stokes scattering does not overlap with fluorescence signals.

**Time-stretch detection of CSRS spectra.** The Stokes scattered photons in the wavelength range from 1,050 to 1,200

nm are coupled into a fiber time-stretcher. We used a standard telecom fiber (SMF-28ultra) for cost-effective time-stretching as it was roughly ten times cheaper than the single-mode fiber for the 1-µm region (e.g., HI1060). However, SMF-28 allows for multi-mode propagation in 1,050-1,200 nm, which causes mode-dependent group delay, preventing one-to-one time-to-wavelength conversion. The higher-order modes are successfully suppressed by connecting a 1-m-long single-mode fiber (HI1060) to the input end of the SMF-28 and adding micro bends at the end of the long fiber. (See Supplementary Note 3 for more details). With this method, a 5-km SMF-28 can stretch the scattered Stokes pulses to 15 ns with a 0.7-dB/km loss. Note that this is the first demonstration of cost-effective time-stretch spectroscopy in the 1-µm region using the telecom SMF-28 fiber, to the best of our knowledge. The total coupling power to the stretching fiber is less than 900 nW. To achieve a high SNR, we utilized a 10-GHz avalanche photodiode (APD) instead of a PIN photodiode. A multiplication factor of 10 of the APD improves the noise-equivalent power (NEP), which is limited by the subsequent trans-impedance amplifier. Signals generated by the APD with 60-ps impulse response are amplified by a 20-GHz low-noise amplifier for efficient use of the dynamic range of a fast oscilloscope with a bandwidth of 16 GHz (WaveMaster 816Zi-B, Teledyne LeCroy). The recorded spectra are processed by numerical low-pass filtering (see Methods and Supplementary Note 6 for details).

**Broadband TS-CSRS spectroscopy of organic molecules.** As a proof-of-concept demonstration, we measured liquid chloroform using our TS-CSRS system. Figure 2a displays a measured temporal intensity, which exhibits repetitive stretched spectra with sharp Raman lines in every 20-ns interval, corresponding to a repetition rate of 50 MHz. Figure 2b shows segmented non-averaged and averaged spectra with spectral distortions due to the non-resonant background (NRB), which is a well-known effect in CARS/CSRS measurements. The time axis is converted to wavelength based on the dispersion of the stretching fiber (See Supplementary Note 4 for more details). A residual peak of the probe pulse at 1,037.5 nm is used for the wavelength calibration. Figure 2c displays the distortion-corrected spectra retrieved by a recently demonstrated machine learning method[23] (See Supplementary Note 5 for more details). Prominent Raman lines of chloroform are clearly observed in the unaveraged spectrum, verifying the broadband CSRS spectroscopy at 50 MSpectra/s. The SNR of the non-averaged spectrum is 27, as evaluated from the temporal intensity data in Fig. 2a. For the SNR calculation, we divided the peak intensity of the Raman line at 668 cm$^{-1}$ by the standard deviation of the noise, which is primarily due to the detector. Figure 2d shows the SNR dependence on the number of averaging, exhibiting the expected square root dependence.

To demonstrate the versatility of our system, we measured various organic molecules and compared them with spontaneous Raman spectra. Figure 3 shows the 10k-averaged spectra of chloroform, dimethyl sulfoxide (DMSO), and acetone compared to the reference spectra obtained with a homemade spontaneous Raman scattering spectrometer. Details of the homemade spontaneous Raman scattering spectrometer are described in Methods. The TS-CSRS spectra show good agreement with the reference spectra.

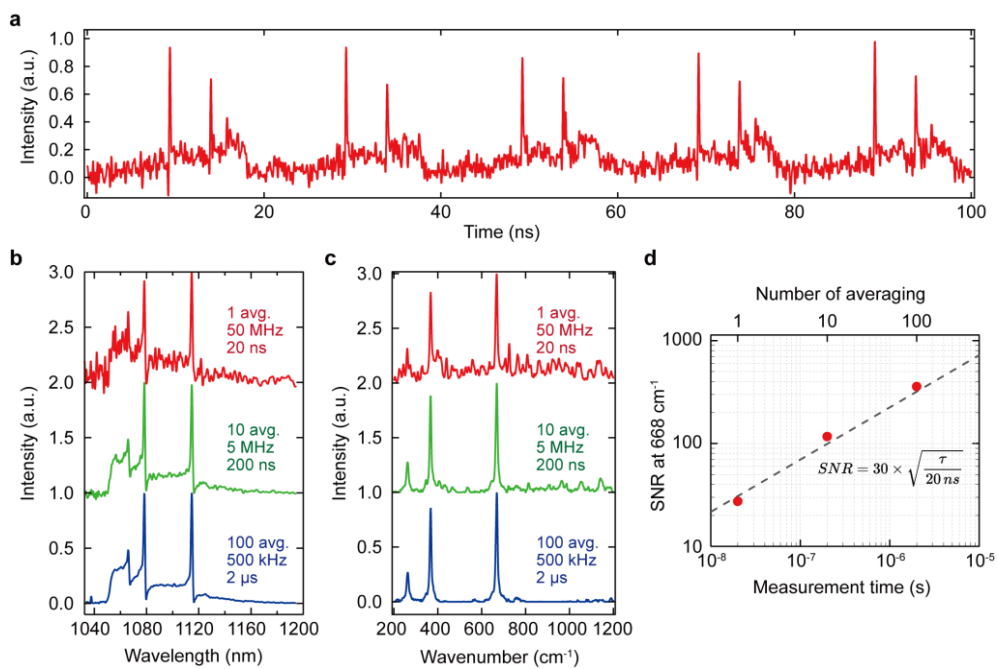

**Fig. 2** Broadband TS-CSRS spectra of chloroform measured at 50 MSpectra/s. **a,** A non-averaged temporal intensity profile of time-stretched pulses. Prominent Raman signal peaks are clearly observed in the single-pulse spectra with a saw-tooth-like envelope repetitively measured every 20 ns. **B,** Segmented non-averaged and averaged spectra. The x-axis is converted from time to wavelength. **C,** Distortion-corrected spectra retrieved by a machine-learning method. The x-axis is converted from wavelength to wavenumber. **d,** SNR of a CSRS peak at 668 cm$^{-1}$ measured with the different number of averaging. The dashed line shows the square root dependence of SNR on measurement time. T: Measurement time (s).

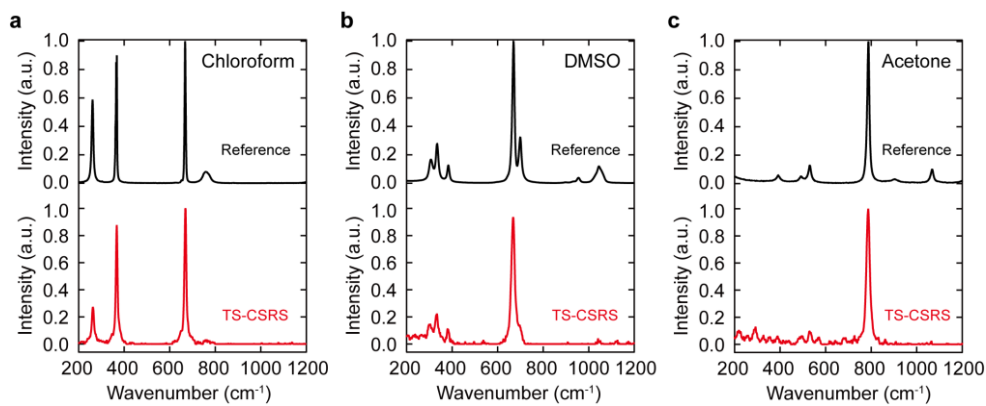

**Fig. 3** Comparison between TS-CSRS and spontaneous Raman scattering spectra of **a,** chloroform, **b,** DMSO, and **c,** acetone. The TS-CSRS spectra are 10k-averaged.

**System characterization.**

The measurable spectral bandwidth can be assessed by a measured spectrum. Figure 4a shows an averaged TS-CSRS spectrum of DMSO before distortion correction. The NRB spectrum spreads over a wavenumber range from 115 to

1,200 cm$^{-1}$, indicating that our TS-CSRS spectrometer can, in principle, cover this spectral region. The cut-on wavelength of the long-pass filter and the excitation pulse duration determine the lower and upper limits of the wavenumber range. As the spectral retrieval process causes an artifact around the edge of the long-pass filtered spectrum, we determine the spectral bandwidth of this system is from 200 to 1,200 cm$^{-1}$.

The spectral resolution of the system is influenced by various factors, such as the impulse response of the detector, the degree of dispersion induced by the stretching fiber, the linewidth of the probe pulse, the sampling bandwidth, and the cutoff frequency of the numerical low-pass filter. Figure 4b displays the estimated spectral resolution of the system. The wavenumber-dependent spectral resolution results from the higher-order dispersion of the stretching fiber. The total combined spectral resolution, including all the above-mentioned factors, is 5-7 cm$^{-1}$ without the 4-GHz numerical low-pass filter and 8-14 cm$^{-1}$ with the filter. The numerical low-pass filter provides a higher SNR by a factor of three at the expense of the spectral resolution.

We measured DMSO diluted with water at different concentrations to assess the signal linearity to the molecular concentration. Figure 4c displays the normalized intensity of a CSRS signal of DMSO at 669 cm$^{-1}$ for various volume concentrations ranging from 0.05 to 0.10, which exhibits a linear relationship, enabling us to determine the minimum detectable concentration. Based on the SNR of the measurements, the minimum detectable volume concentration of DMSO is 0.13, 0.030, and 0.0098 for non-averaged, 10-averaged, and 100-averaged spectra, respectively.

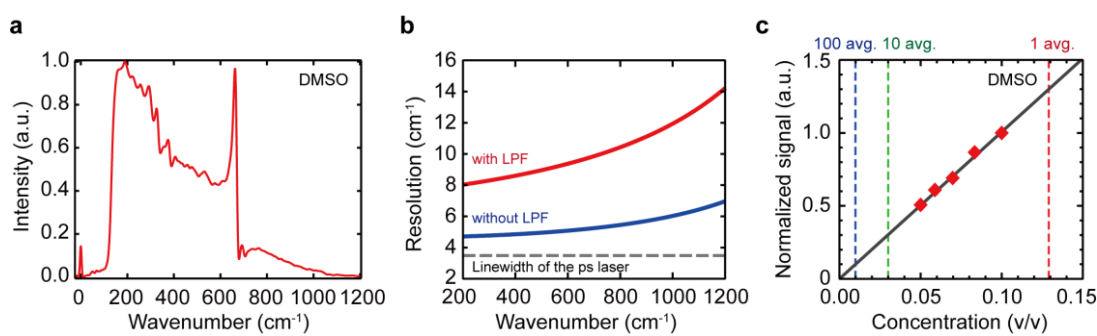

**Fig. 4** System characterization of the TS-CSRS spectrometer. **a,** An averaged TS-CSRS spectrum of DMSO before distortion correction. **b,** Estimated spectral resolutions of the TS-CSRS spectrometer with and without the 4-GHz numerical low-pass filter. The dashed line shows the linewidth of the ps probe pulses. **c,** Signal intensity dependence on the concentration of DMSO diluted with water. The solid black line is a fitting result. The dashed lines represent the detection limit with non-averaged (red), 10-averaged (green), and 100-averaged (blue) spectra.

## Discussions
**Potential system improvement.** We discuss the current limitation and potential improvements in system specifications. First, the spectral measurement rate of 50 MSpectra/s is limited by the repetition rate of the laser system. Therefore, it is possible to make it even faster by using a laser with a higher repetition rate at the expense of spectral bandwidth or resolution under the same detection bandwidth. Second, the measurable spectral bandwidth is mainly limited by the excitation pulse duration. Using sub-10 fs pulses could expand the upper limit of the bandwidth beyond 3,000 cm$^{-1}$, as demonstrated in the previous reports utilizing the ISRS process[24,25]. Finally, the spectral resolution in the current system is optimized for measuring liquid-phase samples, which can be improved by using probe pulses with a narrower linewidth and a longer stretching fiber. Note that there is a trade-off relation among the above-mentioned specifications, such as the measurement rate, bandwidth, and resolution, which is ultimately limited by the detection bandwidth of 10 GHz in the current system. Therefore, higher detection bandwidth promises further improvement in these specifications.

The SNR of the current system is limited by the detector noise. In order to reach the shot-noise limited SNR, a ten-fold increase in signal intensity or a three-fold improvement in the detector's noise equivalent power is required. For the signal intensity enhancement, there is much room for improvement in the focusing condition. In the current system, the focal beam diameter of the probe pulses at the sample is twice as large as the excitation pulses due to the size limitation of the optical components. Therefore, matching the spot sizes could lead to a four-fold improvement in SNR. Using an objective lens with a higher NA instead of the OAPM would also increase the signal intensity. A shorter probe pulse duration could also improve the signal intensity at the expense of the spectral resolution. For example, a pulse duration of 1 ps leads to an 8-fold signal enhancement with a spectral resolution of 14 cm$^{-1}$. Concerning the detector noise improvement, it is challenging to improve the multiplication factor of 10 with 10-GHz APDs. Instead, it would be a solution to use a lower-bandwidth APD with a higher multiplication factor at the expense of the spectral resolution. The selection of a proper detector allows us to take a balance of the trade-off parameters (SNR and resolution) depending on the measurements, providing a great capability to be adopted for multiple purposes.

The wavelength selection of the excitation pulses is an interesting topic for future development. In our demonstration, we used the SHG excitation and near-infrared (NIR) probe pulses, where the two-color CSRS configuration allowed color filtering of the Raman scattered photons. In this configuration, the visible pulse excitation can be used for highly sensitive and selective resonant or pre-resonant Raman scattering measurements with various fluorophors commonly used in bio-imaging[25]. Instead of using the visible SHG pulses, one can also use fundamental NIR fs pulses for impulsive excitation. In this case, a shorter pulse duration with higher pulse energy is available, enabling broader spectral measurements with a higher SNR. In addition, the possibility of sample damage would be reduced due to less probable two-photon absorption processes that cause unfavorable electronic transitions. The Raman scattered NIR photons can be extracted, e.g., by using an orthogonal polarization configuration[6], instead of color filtering.

**Comparison to TS-SRS.** The difference between our TS-CSRS and the previously demonstrated TS-SRS [8] is worth considering. Regarding adaptability to time-stretch spectroscopy, SRS faces a demanding requirement in vertical resolution to distinguish a small Raman gain or loss on top of the large signal. The previous report of TS-SRS showed that data sampling at 12-bit provided higher SNR than 8-bit[8]. However, the bandwidth of the 12-bit oscilloscope was limited to 1 GHz because typical high-speed digitizers are not capable of delivering a high vertical resolution (>12 bits) with a high bandwidth (>10 GHz). Instead, high digitization bits are not required in CARS and CSRS because of their background-free nature. Another interesting difference is the excitation efficiency. TS-SRS relies on single-band excitation with a narrow linewidth ps pulse, while TS-CSRS utilizes fs broadband ISRS excitation. Therefore, under the same excitation pulse energy, the latter provides higher SNR than the former for measuring broadband spectra in the same measurement period.

**Potential applications.** Our high-speed TS-CSRS system expects various potential applications. The continuous measurement capability at 20-ns temporal resolution enables observation of ultra-rapid irreversible complex phenomena such as dynamics of photoreactive proteins[26]. High-speed hyperspectral imaging is another promising direction. With a conventional beam steering system based on resonant/galvanometric scanners, video-rate hyperspectral imaging could be possible. Assuming a 12-kHz resonant scanner, 120 frames/s (8.3 ms per frame) is feasible for measuring 200 x 200 pixels with 10-fold averaging/pixel, which is 10-100 times higher frame rate than the previous state-of-the-art[27–31]. By implementing a faster scanning system, e.g., using 500-kHz piezoelectric transducers[32], even faster imaging at 1,200 frames/s and video-rate 3D imaging could be possible. It also enables large-area imaging of 1,290 x 1,290 pixels at the video rate of 30 frames/s. Our TS-CSRS spectrometer could also advance label-free Raman flow cytometry, which has shown a promising direction in cell biology. The throughput of the state-of-the-art broadband CRS flow cytometer was limited to 2,000 cells/s[12], while those of commercially available fluorescence-based flow cytometers are 10,000-100,000 cells/s. In addition, the low SNR of the previous CRS flow cytometer only measured pre-resonant enhanced signals from carotenoids. Our TS-CSRS could significantly improve the SNR, allowing for measuring a wide range of molecular species at a higher event rate comparable to the fluorescence-based counterparts.

## Conclusion

We developed a TS-CSRS spectrometer and demonstrated broadband CRS spectroscopy spanning over 1,000 cm$^{-1}$ at the record highest rate of 50 MSpectra/s, which was made possible by fully utilizing the high repetition rate of the mode-locked laser. This measurement rate is about 500 times higher than the previous high-speed broadband CRS spectroscopy techniques. The technique has great potential to advance various applications, such as ultrafast measurements of irreversible phenomena, high-speed hyperspectral Raman imaging, and high-throughput Raman flow cytometry, to name a few.

## Methods

**Dispersion management of the fs excitation pulses.** We used OAPMs for focusing and collimating the pulses before and after the SHG crystal. The pulse energy of the SHG output was controlled by changing the polarization of the fundamental input pulses to the BBO crystal to avoid excess dispersion added by optical elements such as neutral density filters. The optical axis of the BBO crystal was fixed horizontally so that the SHG pulses were vertically polarized. To finely tune the amount of dispersion, we adjusted the focal position in the cuvette in the propagation direction.

**Numerical low-pass filtering of CSRS spectra.** A numerical low-pass filter improves the SNR of TS-CSRS spectra. In this work, a 4-GHz low-pass filter was applied to the temporal signal intensity, resulting in 3 times SNR improvement, which was, however, at the expense of degradation of the spectral resolution. Note that we implemented a group-delay-free finite impulse response (FIR) filter, which does not disorder the conversion parameters from time to wavelength. In our experiment with the 4-GHz low-pass filter, the impulse response of the APD of 60 ps becomes 130 ps, resulting in roughly double the resolution. One can select an optimum cutoff frequency of the low-pass filter depending on the samples.

**Homemade spontaneous Raman scattering spectrometer.** We used a homemade spontaneous Raman scattering spectrometer to measure the reference spectra shown in Fig. 3. A Ti:Sapphire continuous-wave laser lasing at 786 nm was irradiated onto the sample with an objective lens. The scattered photons were acquired with the objective and guided into a grating-based spectrometer with a CCD sensor. Volume Bragg gratings were used to reject the Rayleigh scattering. The spectral resolution of the system is 0.77 cm$^{-1}$.

**Funding**
JSPS KAKENHI (20H00125, 21K20500), Research Foundation for Opto-Science and Technology, Nakatani Foundation, UTEC-UTokyo FSI Research Grant Program

**Acknowledgments**
We thank Yoshiki Oishi for taking spontaneous Raman spectra of the samples and Akira Kawai and Makoto Shoshin for commenting on the manuscript.

**Author contributions**
T.I. conceived the concept of the work. T.N. and T.I. designed the system. T.N. constructed the optical systems with the help of K.H.. T.N. performed the experiments and analyzed the data. T.I. supervised the work. T.N. and T.I. wrote the manuscript with inputs from K.H..

**Disclosures**
The authors declare no competing interests.

**Data availability**
The data provided in the manuscript are available from the corresponding author upon reasonable request.

# Supplementary Information for
# Broadband Coherent Raman Scattering Spectroscopy at 50,000,000 spectra/s

Takuma Nakamura[1], Kazuki Hashimoto[1], and Takuro Ideguchi[1,*]

[1] Institute for Photon Science and Technology, The University of Tokyo, Tokyo 113-0033, Japan

*ideguchi@ipst.s.u-tokyo.ac.jp

**Supplementary Note 1: Comparison with other high-speed broadband CRS spectroscopy techniques.**

Table S1 compiles the specifications of representative high-speed CRS spectroscopy techniques in terms of measurement rate, bandwidth, spectral resolution, pulse energy, and SNR. The SNRs are listed not to provide a rigorous comparison but to give a rough evaluation since they strongly depend on the measurement conditions and the samples.

Table S1. Comparison of the specifications of high-speed broadband CRS spectroscopy techniques.

|  | Measurement rate (Spectra/s) | Bandwidth ($cm^{-1}$) | Resolution ($cm^{-1}$) | Pulse energy (nJ) | SNR |
|---|---|---|---|---|---|
| **Rapid-scan FT-CARS[1]** | 50,000 | 1,230 | 4.2 | Pump: 1.5 <br> Probe: 1.75 | N/A |
| **Dual-comb CARS[2]** | 100,000 | ~1,600* | 117 | N/A <br> N/A | 300 |
| **Swept-source SRS[3]** | 111,000 | 650 | N/A | N/A <br> N/A | 34 |
| **Chirped pulse SRS[4]** | 55,000 | 200 | 10.8 | Pump: 0.19 <br> Probe 0.94 | N/A |
| **Time-stretch SRS[5]** | 20,000 | 400 | 10 | Pump: 350 <br> Probe: 1 | ~25* |
| **Time-stretch CSRS (this work)** | 50,000,000 | 1,000 | 8-14 | fs: 1.6 <br> ps: 8.4 | 27 |

*These values are estimated by the authors from the shown data because they are not clearly specified in the referenced papers.

**Supplementary Note 2: Detailed schematic of a TS-CSRS spectrometer**

Figure S1a shows a detailed schematic of a TS-CSRS spectrometer developed in this work. The Spectra of the excitation and probe pulses are presented in Figs S1b and c, respectively.

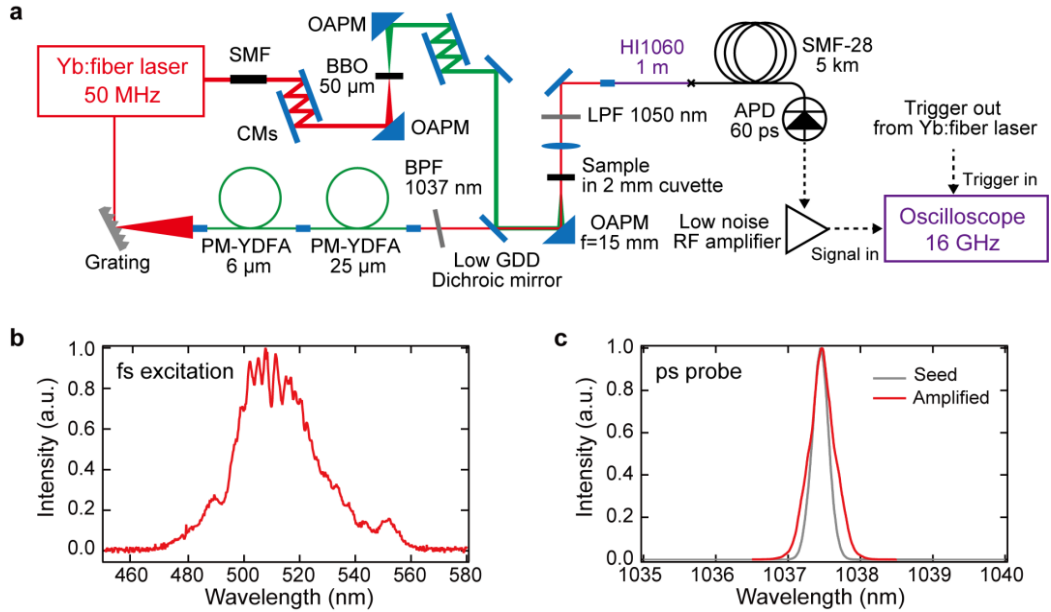

**Fig. S1 a,** A detailed schematic of time-stretch coherent Stokes Raman scattering (TS-CSRS) spectroscopy. PM-YDFA: Polarization-maintaining Yb-doped fiber amplifier, GDD: Group delay dispersion, SMF: single-mode fiber, CMs: chirped mirrors, BBO: β-barium borate, OAPM: Off-axis parabolic mirror, BPF: Band-pass filter, LPF: long-pass filter, APD: avalanche photodiode. **b,** A spectrum of the fs excitation pulse. **c,** Spectra of the ps probe pulse before and after the two-stage amplification.

**Supplementary Note 3: Higher-order-mode rejection and suppression**

As SMF-28 does not perfectly support single-mode propagation in our time-stretching region from 1,050 to 1,200 nm, the non-negligible higher-order modes couple to the fiber. This issue can be solved by connecting a 1-m-long HI1060 fiber to the SMF-28 beforehand. Figure S2 shows temporal intensity profiles of an 8-ps pulse at 1,037.5 nm measured after transmitting a 5-km SMF-28 only (red), the SMF-28 with a 1-m HI 1060 (green), and the SMF-28 with a 1-m HI 1060 and 3-rounds fiber coiling with a quarter inch diameter (blue). As the linewidth of the input pulse is narrow enough, a single short pulse is expected to be observed after the propagation of the 5-km fiber if the fiber supports a single-mode propagation. The measured results clearly show that the SMF-28 produces some satellite pulses caused by different group delays due to the higher-order modes while connecting a 1-m HI1060 fiber and adding fiber coiling suppress the spurious higher-order mode effects. We understand that the former acts as a spatial mode filter, preventing the higher-order modes from coupling into the SMF-28, while the latter introduces a large loss at the higher-order modes. We expect this method to be used for other purposes besides time-stretching, enabling cost-effective single-mode propagation in SMF-28 at wavelengths shorter than 1,250 nm.

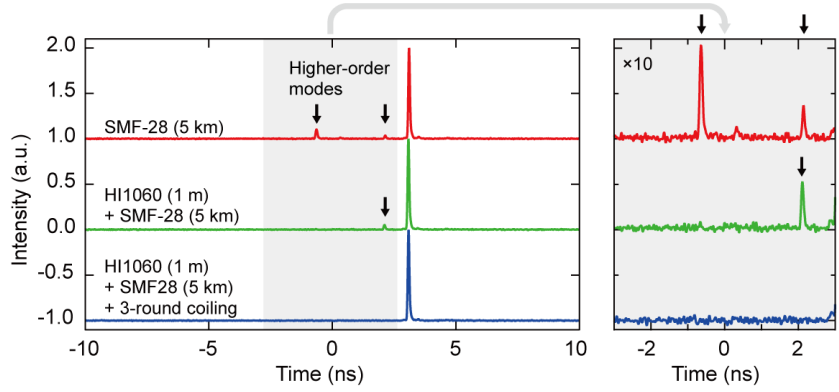

**Fig. S2** Temporal intensity profiles of an 8-ps pulse at 1,037.5 nm measured after transmitting a 5-km SMF-28 only (red), the SMF-28 with a 1-m HI 1060 (green), and the SMF-28 with a 1-m HI 1060 and 3-rounds fiber coiling (blue).

**Supplementary Note 4: Dispersion property of SMF-28 at the 1-μm region.**

The dispersion curve of SMF-28, displayed in Fig. S3a, provided by the manufacturer, is written as

$$D(\lambda) \approx \frac{S_0}{4}\left(\lambda - \frac{\lambda_0^4}{\lambda^3}\right) \qquad (S1)$$

where $D$ is the dispersion of the fiber, $S_0$ is zero dispersion slope of 0.086 ps/(nm·km), $\lambda$ is the wavelength, and $\lambda_0$ is zero dispersion wavelength of 1,313 nm. The datasheet guarantees its validity in a spectral region above 1,200 nm only, which is outside our time-stretching region. Therefore, we experimentally confirmed its validity in the region from 1,050 to 1,200 nm. Figure S3b shows broadband spectra of a supercontinuum light source measured with an optical spectrum analyzer and a time-stretch spectrometer with a 4-km SMF-28. We intentionally add a spectral fringe with an etalon to precisely compare the spectra. The time-to-wavelength conversion for the time-stretch spectrum is made by Equation S1. The fringe periods of the two spectra agree well over the entire range of the spectrum, verifying the dispersion curve also works in the spectral range from 1,050 to 1,200 nm, at least with the shown resolution level. Note that the difference in intensity profile comes from the resolution difference.

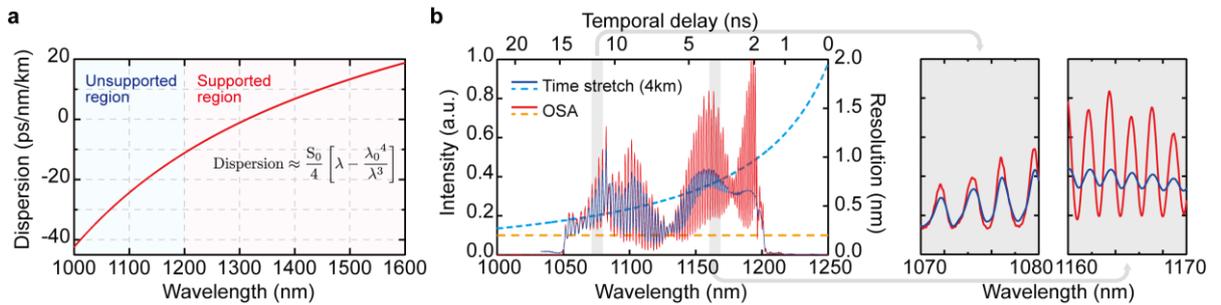

**Fig. S3 a,** Calculated dispersion curve of SMF-28 based on Eq. S1. **b,** Spectrum comparison between time-stretch method with SMF-28 and an optical spectrum analyzer.

**Supplementary Note 5: Spectral distortion correction by a deep-learning method.**

It is known that CARS and CSRS spectroscopy suffer from spectral distortion due to the non-resonant background (NRB). There are various distortion correction methods based on, e.g., time-domain Kramers–Kronig transformation[6], maximum entropy[7], and deep-learning[8,9], where the imaginary part of the third-order nonlinear susceptibility of the molecules is retrieved by removing the NRB. In this work, we applied the recently developed deep-learning method[8]. The procedure of the method, called Specnet, is as follows. 1. Prepare a large number of numerically calculated Raman spectra with random resonant wavenumbers and linewidths, which are the ground truths (answers) for learning. 2. Prepare the same number of random background signals. 3. Make CARS (or CSRS in our case) spectra by taking cross terms of the Raman and the background spectra with a random amount of white noise, which correspond to the inputs for learning. 4. Learn the data sets (inputs and answers) via the convolutional neural network (CNN). More detailed discussions can be found in ref 8. Note that any parameters in Specnet were used with the default values, which worked well with our TS-CSRS results.

**Supplementary Note 6: The effect of a numerical low-pass filter on the impulse response signal.**

Figure S4 shows the impulse response of the detector with and without the 4-GHz numerical low-pass filter. It was measured by detecting an 8-ps probe pulse at 1,037.5 nm, which is sufficiently shorter than the response time of the APD. The measured response time was 60 ps and 130 ps without and with the 4-GHz low-pass filter, respectively. The intensity profiles do not show significant ringing structures. Note that the responses of the 20-GHz amplifier and the 16-GHz oscilloscope are negligible in this measurement.

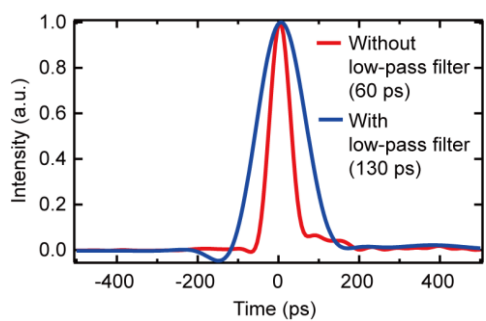

**Fig. S4** Temporal intensity profiles of the impulse response of the APD with and without the 4-GHz numerical low-pass filter.

Figure S5a shows measured TS-CSRS spectra of DMSO with and without the 4-GHz numerical low-pass filter. Figure S5b represents a comparison between the 1k-averaged spectra with and without the low-pass filter, which shows the same spectral shape without additional distortion.

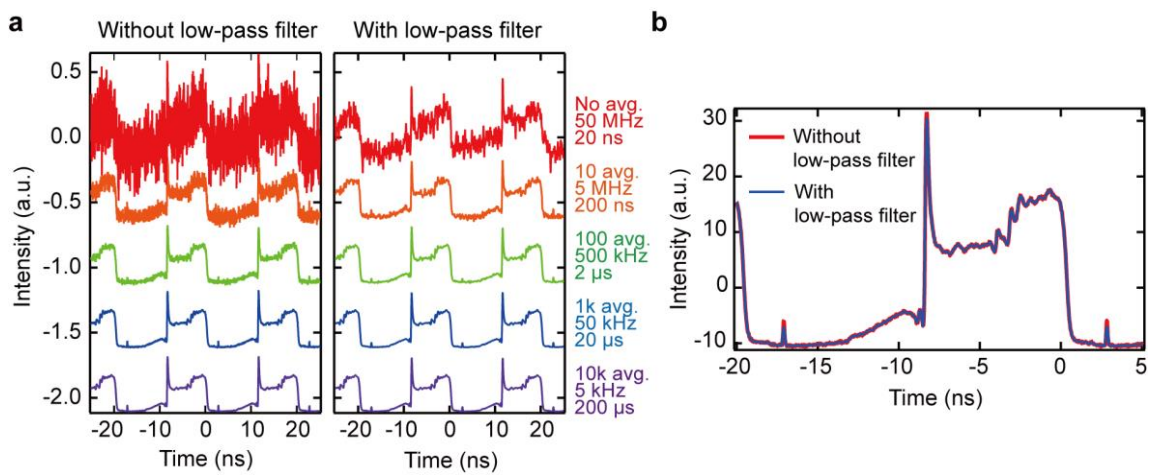

**Fig. S5 a,** TS-CSRS spectra of DMSO measured with and without the 4-GHz low-pass filter. **b,** Comparison between the 1k-averaged spectra.